# Architectural Challenges in Migrating Plan-driven Projects to Agile

Vinod Menon, Roopak Sinha, and Stephen MacDonell
*Auckland University of Technology*
*Auckland, New Zealand*
vinod.menon@datam.co.nz, roopak.sinha@aut.ac.nz, stephen.macdonell@aut.ac.nz

**Abstract**

*Software development has steadily embraced agile software development methodology/method (ASDM) and has been moving away from the plan driven software development methodology (PDM) approaches like waterfall. Given the iterative nature of agile development, the integration of software architecture into the agile way has become challenging. This research identifies the challenges of having a robust architecture in projects already executed by plan driven methods and new projects likewise by conducting a literature review and a case study analysis. The ensuing analysis finds that there are three major areas: people, process and technology, in which these challenges could be mapped.*

**Keywords:** Agile, Waterfall, Plan Driven Methods, Software Architecture, Software Development, Requirements Engineering, Software Development Life Cycle.

## 1. INTRODUCTION

Today the software development process has become synonymous with agile software development methodology/method (ASDM) with more and more practitioners adopting ASDM. On the other hand, software architecture serves as the foundation of software development, and helps build software that is robust, scalable, and low maintenance. However, Agilists perceive that software architecture is deeply rooted in the plan driven methodology (PDM) which gives lot of impetus to "Big up-front design" and heavy documentation (Kruchten, 2010). ASDM favours that we start small and indulge in documentation just as much as needed. In traditional PDM, the architect comes into the project in the beginning and develops an architecture for the project based on the high level requirements or even the detailed requirements as per the system requirement specification document, especially as requirements are not expected to change post signoff. This however does not hold in ASDM as the requirements may keep on changing and evolving through the iterations of the software development life cycle (Lianping & Babar, 2014). These differences make it very challenging to integrate software architecture into the agile way of software development (Abrahamsson, Babar, & Kruchten, 2010). In this paper, we aim to find exactly what these challenges are. The main research question we attempted to answer in this work is:

*"What are the architectural challenges associated while migrating from projects executed by plan driven methods to agile methodology? How real are these challenges?"*

We broke this into the following sub-questions:

**Sub RQ1:** *"Can requirements engineering and software architecture planning go hand in hand in an agile environment?"*

**Sub RQ2:** *"Are there methods to integrate or use software architecture effectively in ASDM?"*

The research methodology involved two aspects: a literature review, and a case study analysis. We carried out a review of 14 contemporary articles on the issue. Some core articles from the earlier agile development years like 2006 were identified to provide inputs to the review. In some of these research articles the researchers have done surveys to obtain the opinion of developers, architects and practitioners of both ASDM and PDM. This helps in providing a more inclusive picture of the contemporary situation with regards to architectural challenges in ASDM. We then analysed a case study of a live enterprise-level HR process first implemented using PDM, and then had to be moved to ASDM. The process was developed in 2005 and has been in use since then. This project was designed to handle employee provisioning and de-provisioning process of the human resources department of a large multinational company head-quartered in India. Over a period of time some functionalities of the process had become redundant. In 2012, the management decided to do a gap analysis to identify modules to be made redundant, revamped, or added some COTS (commercial out of the box software packages) were also to be integrated into the existing project, though the business requirements of the COTS products came from some client stakeholders at a very later stage. Since these changes were to be done in parallel to a live process, it was decided that using ASDM was necessary to quickly realize the benefits of revamping the project. The first author was heavily involved in this process, and the learnings from this experience have helped us make additional distinctions in understanding and answering the research question. Due to confidentiality issues more specific details about the project nature and business objectives cannot be mentioned here. However the challenges faced during the development process can certainly be mentioned.

## 2. RESEARCH ANALYSIS

This section discusses the different challenges that have been identified from the literature review and case study analysis. It is important to note that that software architecture is an essential aspect of any software development process without which the project would lack in the monitoring of quality attributes and finally cease to give desired results. To enable an effective software architecture for a project it must be documented and updated regularly with the changes in the project. In PDMs like water-fall the architecture design is done in the initial stages of the project with help of the high level requirements and the expertise of the architect (Diego & Martín, 2013). The software architecture deliverable is an architecture document which is essentially a cross cutting document of the 4+1 view-sets as per Philip Kruchten (2010), focusing on communication, quality, design patterns and the hardware and software stack. The architecture document tries to address different stakeholders like the business, the project managers, the business analysts, the development and testing team, the system engineers. The architecture document maps the logical, process, physical, developmental and scenario based view-sets in the design.

Producing an architecture document is quite easy with the traditional PDMs, but when the project is executed using ASDM like Scrum or XP, it becomes challenging. One reason for this is mainly because ASDMs start projects with minimum information and do incremental software development in small iterations called sprints (Falessi, Cantone, Sarcia, Calavaro, Subiaco, & D'Amore, 2010). In this research some major challenges have been identified and grouped into categories as people, process, technological and requirements engineering challenges. This classification of challenges arises from the authors' survey of available literature and further research is required to ensure that it is complete.

### 2.1 People Factors

People factors have emerged as some of the major challenges from the literature review and the case study analysis and are mentioned in **Table 1**.

**Table 1** shows that perception issues of Agilists considering software architecture as a big upfront design strategy with its roots in plan driven methods is one of the major influencers. Minor factors include the differences between communication, collaboration and negotiation strategies used in PDM and ASDM, as well as the team's understanding of the role of software architecture in an agile environment. The product owner's approach towards non-functional requirements is a major factor which can increase the technical debt. A team's overall appreciation of quality attributes may also contribute to the challenges in implementing software architecture within agile projects. The architect's role also undergoes a major change in ASDM. The time commitment that an architect has to give in ASDM is far greater than the consultative approach taken in PDM. In agile architecture the architect needs to be present in almost all the sprints to map the software architecture effectively, besides negotiating on a regular basis with the product manager on the architectural backlog.

Table 1: People Factors.

| People Factors | Degree of Influence | |
|---|---|---|
| | Major | Minor |
| Communication, collaboration, negotiation | | ✓ |
| Understanding the role of software architecture in Agile environment | | ✓ |
| Product Owner's approach towards non-functional requirements | ✓ | |
| Agile team's appreciation of quality attributes | ✓ | |
| Perception issue of considering software architecture as big upfront design strategy from PDM. | ✓ | |

Table 2: Process Factors.

| Process Factors | Degree of Influence | |
|---|---|---|
| | Major | Minor |
| Integrating software architecture in small projects and enterprise wide large projects | | ✓ |
| software architecture integration with the sprint requirement engineering process | ✓ | |
| Mapping the non-functional requirements in the Agile software development life cycle | ✓ | |
| Software architecture's association with refactoring in Agile software development life cycle (Breivold, Sundmark, Wallin, & Larsson, 2010) | ✓ | |
| Developing the process view given the incremental development in the agile development process | ✓ | |

### 2.2 Process Factors

ASDM is quite different from PDM in terms of process. Software architecture was never invented to work in ASDM and hence Agilists always view software architecture to be a legacy document- intensive approach. This introduces the challenge of adopting software architecture to the process of ASDM. This research has been able to identify some major process challenges as mentioned in **Table 2**.

Process factors like integration of software architecture in the agile requirements engineering process (Philip, Afolabi, Adeniran, Ishaya, & Oluwatolani, 2010), mapping of non-functional requirements, learning from refactoring in-order to contribute to the architectural document and producing a process view with the help of UML activity diagrams have major impact on the architectural decisions made (Breivold et al., 2010).

As per Philip et al. (2010), software architecture can be integrated into all phases of an agile software development life cycle. If the architect fails to draw the attention of the stakeholders in each of these aspects throughout the ASDM process, the resulting architecture would be compromised in terms of quality.

### 2.3 Technology Factors

Agile architecture is incremental in nature as the architect obtains information from the developers in every sprint which the team learns during the coding and refactoring process. Though the agile architect starts the first sprint with some upfront design, the challenge is that the architectural information only flows in incrementally over subsequent sprints as new or more concrete requirements emerge. This makes the task of mapping the technology stack a longer process as compared to PDMs. Some of the technological challenges associated with agile architecture both in case of new projects or executing agile projects that have been earlier done by using PDMs identified in this research are as mentioned below:

- Framing the development view
- Decomposition of requirements to develop the logical view (Madison, 2010)
- Framing the physical view
- Scalability
- Integrating the Architectural backlog with the Product backlog
- Integration of COTS software and hardware packages (Nuseibeh, 2001)

If there is a new module to be included in a project previously executed using PDM then the existing architecture can be used, but this depends on the compatibility between software and hardware stacks used then and now. There is also an architectural challenge with respect to integrating COTS into a legacy software system (Boehm & Turner, 2003; Nuseibeh, 2001). The challenge of scalability is another major aspect in terms of technology factors. All six factors listed above have a ***major*** impact on the architecture document which constitutes developing the logical, developmental and physical view from a system engineer's perspective. It should be noted that, like in the case study under consideration, if the technology stack of the new modules to be added into the existing project is different from that of the legacy system, the existing architecture will be impacted and may require updating. It is quite possible that a COTS product with a new set of business requirements is added to an existing project. While this is expected in agile environments, the architect must make provisions to accommodate such changes which could add to the architectural backlog of the project (Nuseibeh, 2001). Our literature review also finds that the architectural backlog is an area which must receive similar attention as the product backlog. This can be achieved by making use of architectural use-cases and negotiating with the product owner in order to reduce technical debt (Madison, 2010). If care is not taken, the cost of refactoring and fixing bugs would go up exponentially (Miyachi, 2011). If these technical factors are considered then a robust architecture design can be created.

### 2.4 Requirements Engineering in Sprints and Software Architecture

Requirements engineering caters to gathering two types of requirements. Functional requirements describe specific functions or behaviours of a system whereas non-functional requirements (NFRs) describe criteria used to judge the operation of a system. Software architecture tends to be more focussed on NFRs and associated quality attributes. If a development team ignores NFRs, the project may suffer from huge refactoring costs and decreasing sprint velocity, eventually resulting in the project deviating from the baseline requirements, not being delivered in time, and the maintenance of the project being compromised (Abrahamsson et al., 2010). At the architectural level, NFRs are quality attributes of a project which can be further classified into two types as below:

A. **Executable qualities** – Attributes that have a direct impact on the project health. E.g. usability and security (Abrahamsson et al., 2010).
B. **Operational qualities** or **evolution qualities**- The impact of these attributes can be seen over a period of time even when the project is in a post-production life cycle. E.g., scalability, extensibility, testability, integration with COTS and maintainability (Falessi et al., 2010).

According to the twin peak theory by Bashar Nuseibeh (2001), functional requirements in an ASDM are elicited in an iterative style and an architectural design gradually evolves from these iterative sprints of requirements engineering (Cleland-Huang, Hanmer, Supakkul, & Mirakhorli, 2013). Early elicitation of prominent user stories that contribute to the architectural design gives both the development team and the architect a clear vision of the architectural roadmap (Fraser, Hadar, Hadar, Mancl, Miller, & Opdyke, 2009). This also requires the team to appreciate the architectural benefits that could be reaped like avoiding painful refactoring and having an incremental sprint velocity with early realization of business functionality (Cleland-Huang et al., 2013). According to Madison (2010), there are various interaction points like upfront planning, storyboarding and sprints in the agile software development life cycle where the architect must participate in the requirements engineering process to gather information on the architectural aspects.

## 3. DISCUSSION AND FINDINGS

### 3.1 Main Research Question

Our main research question was *"What are the architectural challenges associated while migrating from projects executed by plan driven methods to agile methodology? How real are these challenges?"*

The main findings of this research as discussed in the analysis are the architectural challenges like people, process, technology and requirement engineering aspects arising out the iterative nature of ASDM to produce viable robust software. Another major finding is the new dynamics of the architect's role as a team member. This answers part of our main research question, however the second part of the research question is yet to be answered. According to Abrahamsson et al. (2010), ASDM was never meant to perform the way PDM performed software development. So if carefully observed it is necessary to find ways to integrate software architecture into the way ASDM functions. According to Falessi et al., 2010, there are many reasons why software architecture is required for ASDM. In a decreasing order of relevance, the most important aspects are communication of architectural uses cases across all stages of the software development life cycle, support in system design and development, documentation of risks and assumptions, providing alternative solutions and design patterns, effective communication of functionality to stakeholders, effective evaluation and analysis, and enabling transition to or integration of new software architecture to legacy software architecture.

### 3.2 Sub Research Questions

**Sub RQ1:** *Can requirements engineering and software architecture planning go hand in hand in an agile environment?*

The other findings based on the literature review and the case study analysis are that the requirements engineering and software architecture must go hand in hand. As per to Madison (2010), software architecture has certain interaction points where it interacts with agile software development process. These interaction points help the architect to obtain valuable information on the NFRs. Besides these a quality workshop as proposed by Nord & Tomyako (2006) would also enable the architect to map the quality attributes of the NFRs as discussed in section 5.1.4. Though in ASDM process both the functional and non-functional requirements evolve over a number of iterations/sprints, it is quite possible to create incremental architectural design over these iterations. This however would require the architect to participate in up-front planning, storyboarding and the sprints, so the scope of the architect's work in ASDM increases considerably. In ASDMs the architect's role is participative, being more of a team member, mentor or a guide for the team in terms of non-functional requirements. In PDMs the architect's role is more consultative. Up-front planning and start-up documentation is a must which the architect would have to do in both ASDM and PDM. Thus requirements engineering and software architecture do go hand in hand as is found in this research.

**Sub RQ2:** *Are there methods to integrate or use software architecture effectively in ASDM?*

This research also identifies methods to integrate software architecture into ASDM based on the efforts of Software Engineering Institute (SEI) of Carnegie Mellon University (Nord & Tomayko, 2006) . These methods like the Attribute driven design method, Architecture Trade-off Analysis Method and Cost- Benefit Analysis Method enable consistency between the agile implementation and architectural design (Nord & Tomayko, 2006).

### 3.3 Recommendations

Inferences and recommendations of this research analysis are as follows:

- Architectural deliverables must be given equal priority along with business requirements.

- Early up-front architectural design is a must.

- Clarity on when to freeze architectural requirements is required.

- **Role of the architect:** A successful agile architect must have the following skill-sets:

    - Clear understanding of agile practices

    - Interact with the development team to gather architectural insights

    - Must be able to guide and mentor the team on non-functional requirements

    - Must be able to negotiate with the product owner on the architectural backlog visa/vis the product backlog

    - Strong stakeholder communication skills to highlight the trade-offs in the non- conformance of architectural standards.

**Future work**: While this research has provided an initial insight into the way agile architecture functions, it is limited in scope as the data collection has been limited to only a literature review of contemporary articles and one case study. Future work would involve further case studies and a survey of practitioners of agile and software architects.

## 4. CONCLUSIONS

This research brings to fore the main aspect that although there are challenges in integrating software architecture with ASDM, it is necessary to integrate them. The main categories of challenges identified by the research are people, process, technology and requirements engineering factors. Requirements engineering was found to aid the integration process of software architecture into ASDM. Some factors like using architectural use-cases, and negotiation with the product owner regarding the architectural backlog have been identified to deliver on architectural design. Thus the research effort addresses the main question of the architectural challenges faced while moving from projects executed in PDMs to ASDMs and also examines the perception issues associated with these challenges.